\documentclass[12pt]{article}
\usepackage{graphicx}
\usepackage{wrapfig}
\usepackage{cite} 
\pdfoutput=1 

\newcommand\pubnumber{CMS CR-2016/420}
\newcommand\pubdate{\today}

\def\institute{H.H. Will Physics Laboratory, University of Bristol\\
Tyndall Avenue, Bristol, BS8 1TL, UK\\
and\\
Interuniversity Institute for Higher Energies, Vrije Universiteit Brussel\\
Pleinlaan 2, 1050 Brussels, Belgium}
\def\support{\footnote{Work supported by the Science and Technology Funding Council\\
and the Interuniversity Institute for Higher Energies.}}

\def\Title#1{\begin{center} {\Large #1 } \end{center}}
\def\Author#1{\begin{center}{ \sc #1} \end{center}}
\def\Address#1{\begin{center}{ \it #1} \end{center}}

\newcommand\pubblock{\rightline{\begin{tabular}{l} \pubnumber\\
         \pubdate  \end{tabular}}}
\newenvironment{Abstract}{\begin{quotation}  }{\end{quotation}}
\newenvironment{Presented}{\begin{quotation} \begin{center} 
             PRESENTED AT\end{center}\bigskip 
      \begin{center}\begin{large}}{\end{large}\end{center} \end{quotation}}

\def\beq{\begin{equation}}
\def\eeq#1{\label{#1}\end{equation}}
\def\eeqn{\end{equation}}

\def\beqa{\begin{eqnarray}}
\def\eeqa#1{\label{#1}\end{eqnarray}}
\def\eeqan{\end{eqnarray}}

\let\bar=\overbar

\def\Dslash{\not{\hbox{\kern-4pt $D$}}}
\def\dslash{\not{\hbox{\kern-2pt $\del$}}}

\def\msb{{\bar{\ssstyle M \kern -1pt S}}}

\def\ttbar{\ensuremath{\rm{t}\overline{\rm{t}}}}
\def\tttt{\ensuremath{\rm{t}\overline{\rm{t}}\rm{t}\overline{\rm{t}}}}

\def\sigmattttSM{\ensuremath{\sigma_{\tttt}^{\rm{SM}}}}

\def\CLS{\ensuremath{\rm{CL}_{\rm{S}}}}

\begin{document}
\begin{titlepage}
\pubblock

\vfill
\Title{Search for the production of four top quarks at the CMS experiment at $\sqrt{s}=13$~TeV}
\vfill
\Author{Lana Beck\support~on behalf of the CMS collaboration}
\Address{\institute}
\vfill

\begin{Abstract}
The production of four top quarks at the LHC is an incredibly rare process with an overwhelmingly large background. The cross section can be enhanced in several models of new physics which makes it a particularly interesting process to study. The search in the single lepton and the opposite-sign dilepton channels at the CMS experiment at $\sqrt{s}=13$~TeV is presented. Multivariate algorithms are employed to enhance the separation of the signal and background processes. A template fit is used to set limits on the standard model cross section, combined across both channels.
\end{Abstract}

\vfill
\begin{Presented}
$9^{th}$ International Workshop on Top Quark Physics\\
Olomouc, Czech Republic,  September 19--23, 2016
\end{Presented}
\vfill
\end{titlepage}
\def\thefootnote{\fnsymbol{footnote}}
\setcounter{footnote}{0}

\section{Introduction}

The predominant production of top quarks at the LHC is via pair production into a top and anti-top pair which has a cross section of 831~pb at $\sqrt{s}=13$~TeV~\cite{Topplus}. Single top quark production is much rarer but has also been observed~\cite{singleT}.\begin{wrapfigure}{r}{0.4\textwidth}
\includegraphics[width=0.55\textwidth]{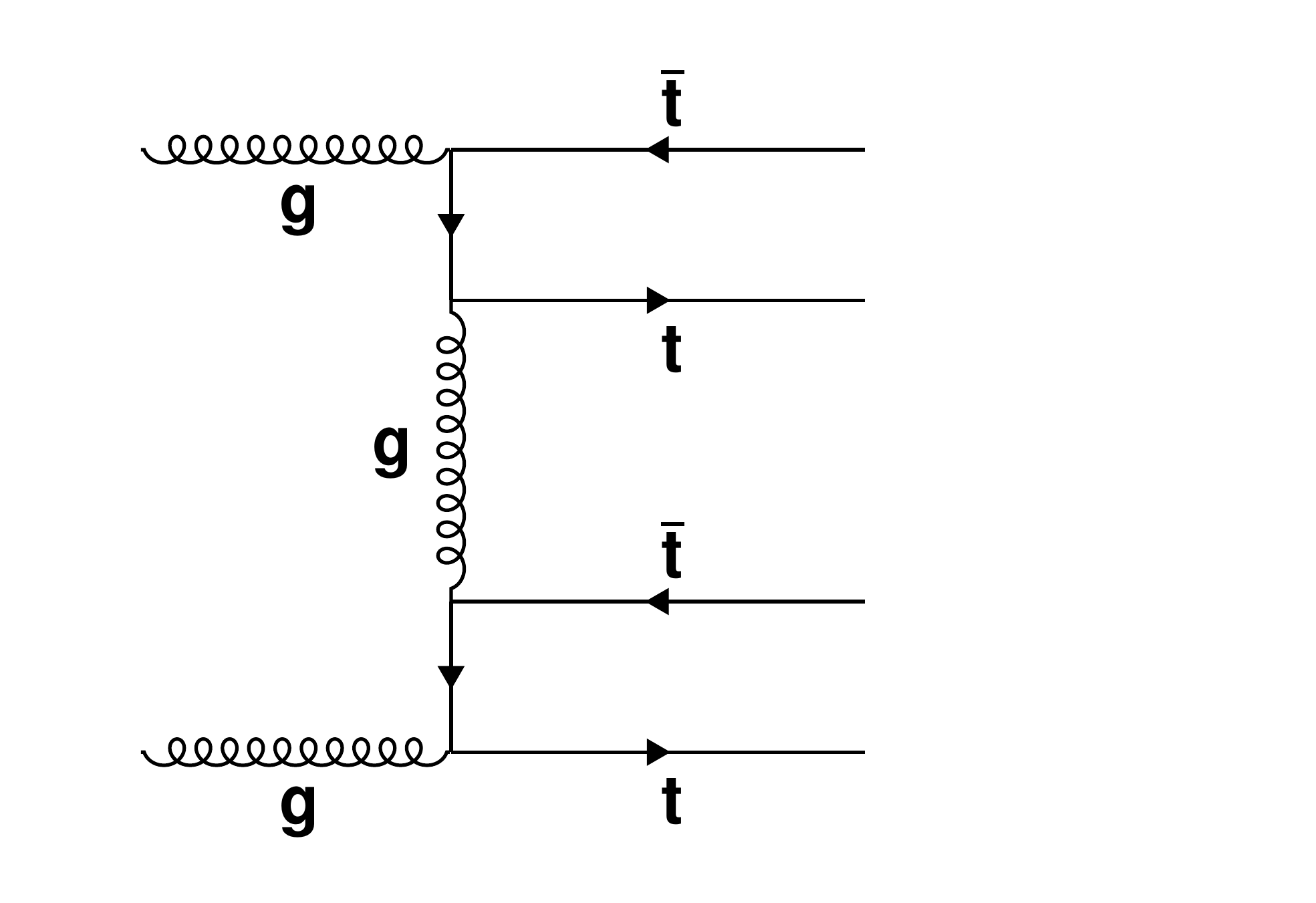}
\caption{The dominant diagram for the production of four top quarks via gluon fusion.}
\label{fig:tttt}
\end{wrapfigure}
The production of four top quarks is an extremely rare process in the standard model (SM) with a cross section of 9.2~fb at $\sqrt{s}=13$~TeV~\cite{xsec}.
This cross section can be enhanced in many beyond the standard model (BSM) scenarios. BSM models which can produce final states with four top quarks include supersymmetry, universal extra dimensions, models with massive coloured bosons and model with Higgs or top quark compositeness~\cite{CompScalars,dm,heavyS,multitop,yukawa,Zprime}. The dominant production mechanism via gluon fusion is shown in Fig.~\ref{fig:tttt}. 
Final states are defined by the decay of the top quark where it is assumed that the top quark decays into a W boson and a bottom quark $100\%$ of the time. The W boson can then either decay to a charged lepton and a neutrino or to a quark and anti-quark which will produce two jets in the detector in addition to the bottom quark jet. The single lepton channel contributes the largest branching fraction ($39\%)$ whilst the dilepton channel has the second largest branching fraction ($30\%$), which is the reason these two channels are chosen to study. Only final states containing electrons and muons are considered.
Limits have been placed on the production cross section of four top quarks at $\sqrt{s}=8$~TeV by CMS of 32 fb observed and 32$\pm$17 expected~\cite{papereight} where the predicted cross section is 1.3~fb~\cite{xsec}.
	
\section{Analysis}

This analysis was performed on the full 2015 dataset collected by CMS~\cite{CMS} which corresponds to 2.6~fb$^{-1}$ of data. Triggers were used based on the presence of a single muon (electron) with a $p_{\rm{T}}\geq26~(30)$~GeV in the single lepton channel. The dilepton channel used triggers which required the presence of an electron or muon with $p_{\rm{T}}\geq17$~GeV in combination with a second lepton of $p_{\rm{T}}\geq12~(8)$~GeV for an electron (muon). The single lepton (dilepton) channel is required to have exclusively one (two) leptons to keep the channels orthogonal to each other. In the dilepton channel only events containing an opposite-sign dilepton pair are studied. Six (four) total jets are required in the single lepton (dilepton) channel and both channels require at least two of these jets to be identified as b-quark jets.
Additionally a Z veto is applied in a 30~GeV window of the Z mass in same-flavour dilepton events. A requirement on the sum of the transverse hadronic energy of $\rm{H}_{\rm{T}}>500$~GeV is applied in the dilepton channel as it enhanced sensitivity to the signal.

\section{Hadronic top quark reconstruction}

The identification of hadronic top quarks can help to distinguish between four-top-quark production where there are three (two) hadronic tops in the final state of the single lepton (dilepton) channel compared to one (zero) in the \ttbar~final state. There are many combinations of three jets possible in these high jet multiplicity selections which motivates the use of multivariate analysis to identify those jets originating from a top quark. Variables such as the invariant mass of the trijet and the invariant mass of dijet with the smallest $\Delta$R separation are used as input variables to a Boosted Decision Tree algorithm. The BDT$_{trijet}$ is trained on \ttbar~events and evaluated on all datasets. The first highest-ranked trijet, BDT$_{trijet1}$, can be used as a discriminating variable in the dilepton channel. In the single lepton channel the jets from the first highest-ranked trijet are excluded from the collection of jets and the next highest ranked trijet, BDT$_{trijet2}$, is used as a discriminating variable in the single lepton channel, as semi-leptonic \ttbar~may have a reconstructible hadronic top.


\section{Multivariate analysis using Boosted Decision Trees}
\begin{figure}[ht!]
\centering
\includegraphics[width=0.4\textwidth]{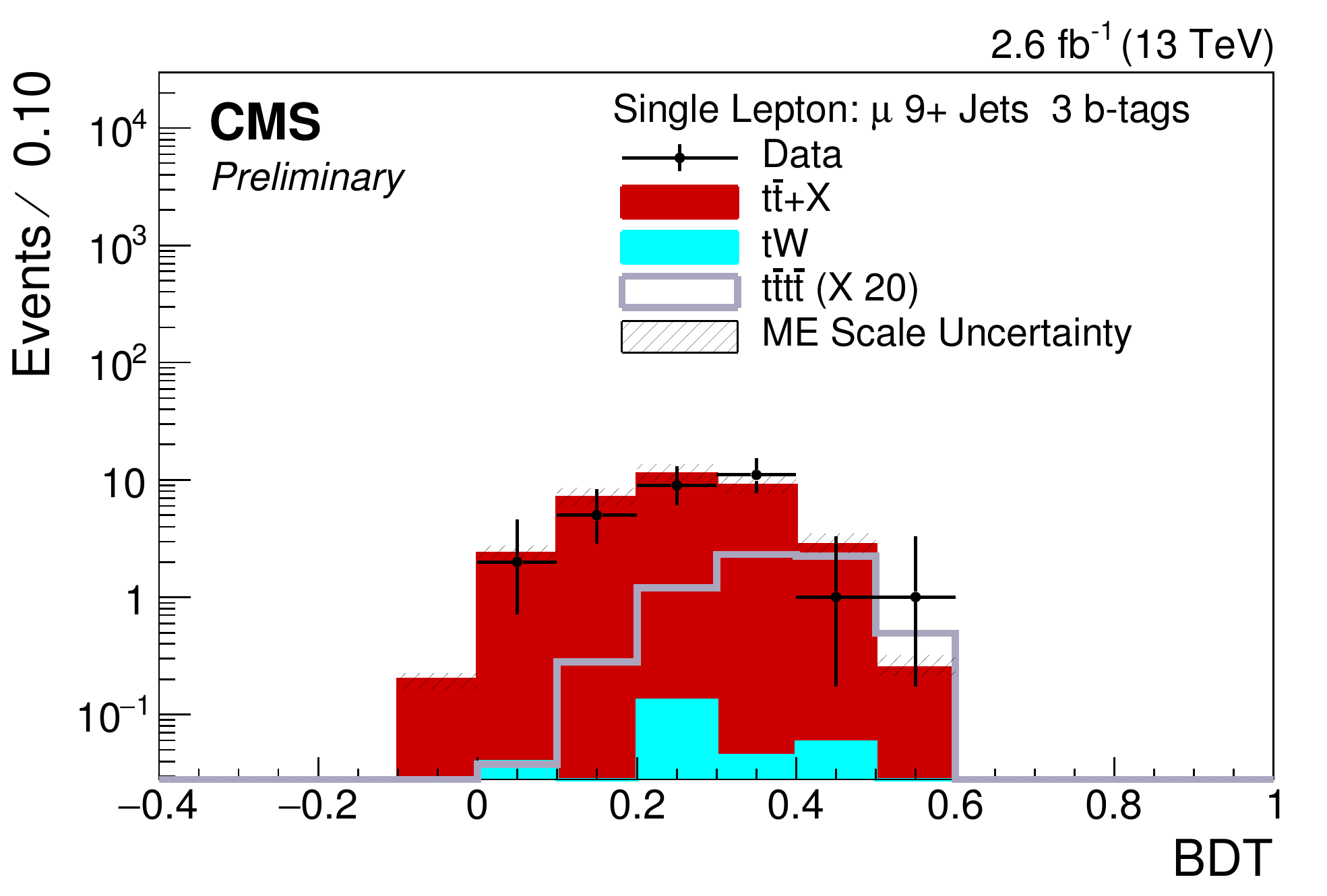}
\includegraphics[width=0.4\textwidth]{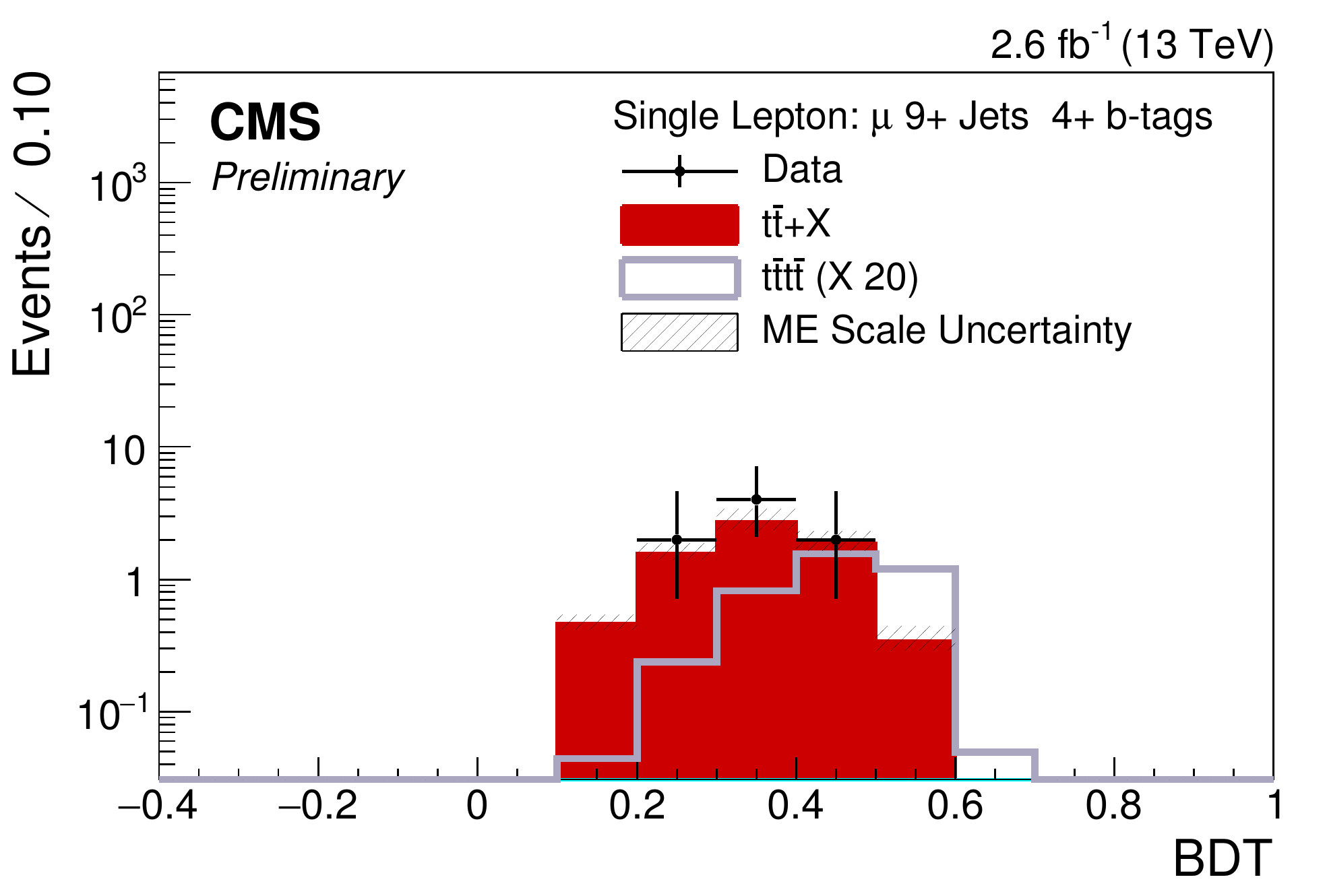}
\includegraphics[width=0.4\textwidth]{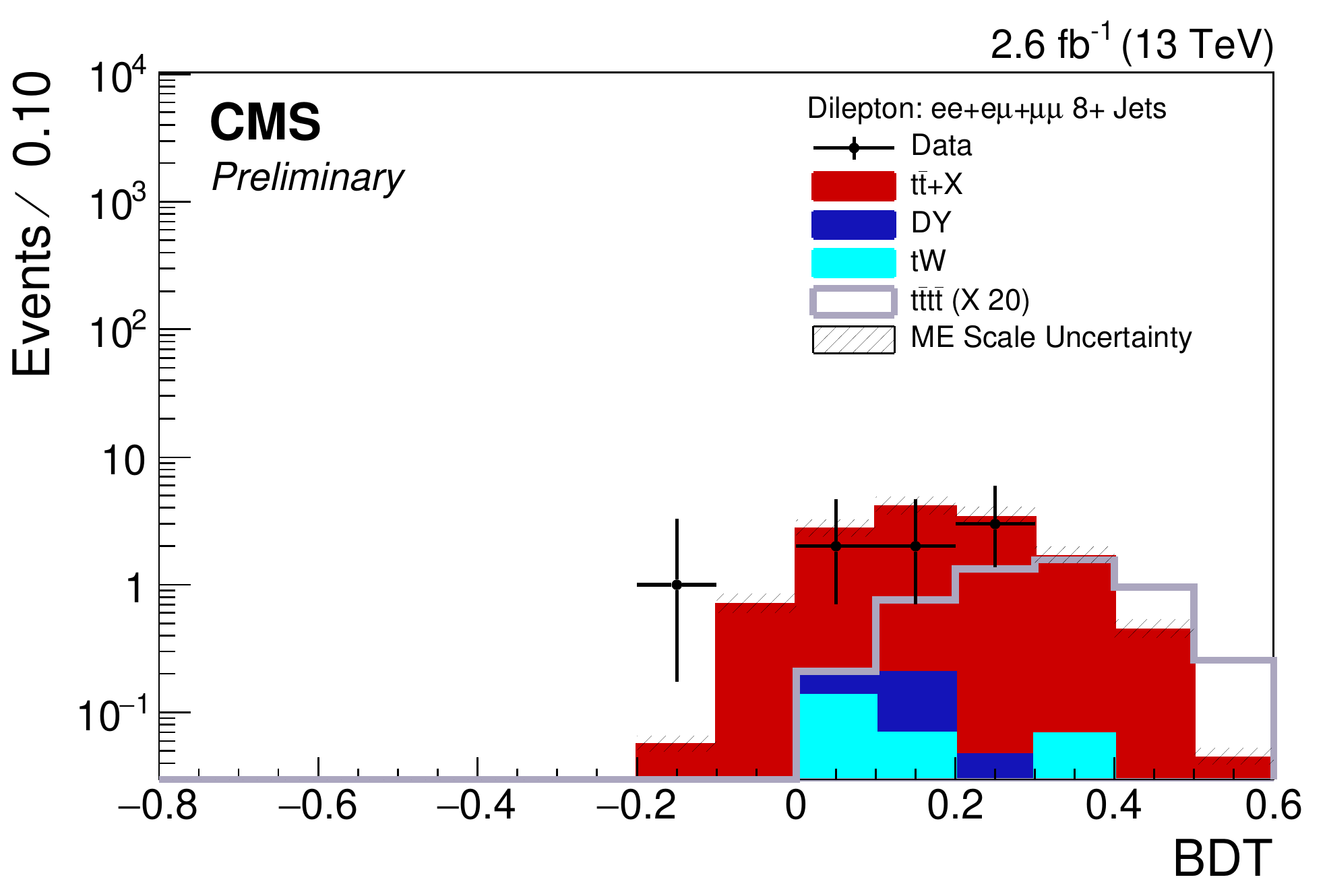}
\caption{BDT output distributions in the $\mu$ + jets channel for the $\geq9$ jets and $3$ b-tagged jets~(top-left) $\geq9$ jets and $\geq4$ b-tagged jets~(top-right) categories and for the dilepton channels summed across lepton flavours for $\geq$8 jets (bottom)}
\label{fig:BDT}
\end{figure}

Further to the hadronic top reconstruction another BDT, BDT$_{event}$, is used to increase the separation between the signal \tttt~events and the background \ttbar~events, using BDT$_{trijet1}$ and BDT$_{trijet2}$ as input variables in the dilepton and single lepton channels respectively. Other variables, which are based on the b-jet content, event topology and event activity, are provided as input variables to BDT$_{event}$. The output histograms are split into categories according to the number of jets in the events. The categories are 6, 7, 8, $\geq$9 jets in the single lepton channel and 4-5, 6-7 and $\geq$8 jets in the dilepton channel. The single lepton channel, which is less statistically limited, is further categorised according to the number of b-tagged jets in the event; 2, 3, $\geq$4 b-tagged jets. The higher jet and b-tagged jet categories are most sensitive to the signal whereas the lower jet and b-tagged jet categories act as control regions which constrain the background. 	Some of most sensitive categories are shown in Fig.~\ref{fig:BDT} where it can be seen that the \tttt~signal is found at higher BDT values on average compared to the main \ttbar~background.

\section{Limit setting and summary}
A combined maximum likelihood fit is performed in the single lepton and dilepton channels. Limits are extracted using the asymptotic \CLS~method and are shown in Table~\ref{tab:limits_single}. It can be seen that the single lepton channel is more sensitive. The combined result of a 95\% CL upper limit on the production cross section of \tttt~is 10.2~$\times$~\sigmattttSM~observed with $10.8^{+6.7}_{-3.8}~\times$~\sigmattttSM~expected which corresponds to 92.8 fb observed and $99.4^{+61.6}_{-35.0}$ expected~\cite{pas}.

\begin{table}[ht!]
	\centering
		\vspace{20pt}

	\begin{tabular}{ l | c c c |  c }
		Channel  & \multicolumn{3}{c|}{Expected Limit (x \sigmattttSM~) } & Observed Limit (x \sigmattttSM~) \\	
		\hline 
		\rule{0pt}{4ex} 

                Single Lepton  & \multicolumn{3}{c|}{$12.7^{+7.8}_{-4.4}$} & $16.1$   \\ 
		\rule{0pt}{4ex} 
                Dilepton  & \multicolumn{3}{c|}{$22.3^{+16.2}_{-8.4}$} & $14.9$   \\
		\rule{0pt}{4ex} 
                Combined  & \multicolumn{3}{c|}{$10.8^{+6.7}_{-3.8}$} & $10.2$                 \vspace{3pt}  \\
	\end{tabular}
		\caption{Expected and observed 95\% CL upper limits on the standard model four top quark production as a multiple of \sigmattttSM~. The values quoted on the expected limits are the $1\sigma$ uncertainties.}	
	\label{tab:limits_single}
  \end{table}

\end{document}